# DRUG REPURPOSING FOR PARKINSON'S DISEASE USING RANDOM WALK WITH RESTART ALGORITHM AND THE PARKINSON'S DISEASE ONTOLOGY DATABASE

Pratham Kankariya[1][1], Rachita Rode[1][2] and Kevin Mudaliar[1][3]

[1] *De*partment of Electronics and Telecommunication*, K. J. Somaiya Institute of Technology*
[2]*Mumbai University, Sion, Mumbai, 400022, Maharashtra, India*
p.kankariya@somaiya.edu , rachita.rode@somaiya.edu, kevin@mudaliar@somaiya.edu

Abstract: Parkinson's disease is a progressive and slowly developing neurodegenerative disease, characterized by dopaminergic neuron loss in the substantia nigra region of the brain. Despite extensive research by scientists, there is not yet a cure to this problem and the available therapies mainly help to reduce some of the Parkinson's symptoms. Drug repurposing (that is, the process of finding new uses for existing drugs) receives more appraisals as an efficient way that allows for reducing the time, resources, and risks associated with the development of new drugs. In this research, we design a novel computational platform that integrates gene expression data, biological networks, and PDOD database to identify possible drug repositioning agents for PD therapy. By using machine learning approaches like the RWR algorithm and PDOD scoring system we arrange drug-disease conversions and sort our potential sandboxes according to their possible efficacy. We propose gene expression analysis, network prioritization, and drug target data analysis to arrive at a comprehensive evaluation of drug repurposing chances. Our study results highlight such therapies as promising drug candidates to conduct further research on PD treatment. We also provide the rationale for promising drug repurposing ideas by using various sources of data and computational approaches.

## 1 INTRODUCTION

Drug Parkinson's disease (PD) is the second most common neurodegenerative disorder, affecting approximately 1% of individuals over the age of 60 [1]. The primary pathological hallmark of PD is the progressive degeneration of dopaminergic neurons in the substantia nigra region of the midbrain, leading to a deficiency in dopamine production [2]. This dopamine deficiency results in various motor symptoms, including tremors, rigidity, bradykinesia, and postural instability, as well as non-motor symptoms such as cognitive impairment, sleep disturbances, and autonomic dysfunction [3].

While current treatments for PD, such as levodopa and dopamine agonists, can provide symptomatic relief, they do not address the underlying cause of the disease or halt its progression [4]. Moreover, long-term use of these medications is often associated with adverse effects, such as dyskinesias and motor fluctuations, highlighting the need for more effective and safer therapeutic approaches [5].

Drug repurposing, also known as drug repositioning or reprofiling, has emerged as a promising strategy in drug development, particularly for complex diseases like PD [6]. This approach involves identifying new therapeutic applications for existing drugs that have already undergone extensive safety and pharmacokinetic testing, potentially reducing the time and cost associated with traditional drug discovery processes [7].

---

[1] https://www.linkedin.com/in/pratham-kankariya

[2] https://orcid.org/0000-0000-0000-0000

[3] https://orcid.org/0000-0000-0000-0000

In recent years, the integration of computational methods, such as machine learning and network-based approaches, with diverse biological data sources has facilitated the identification of potential drug repurposing candidates [8]. One such data source is the Parkinson's Disease Ontology Database (PDOD), which curates and integrates various types of information relevant to PD, including gene expression data, biological pathways, and drug-target interactions [9].

The objective of this study is to develop a computational framework that leverages the PDOD, gene expression data, and biological networks to identify and prioritize existing drugs for potential repurposing in PD treatment. By integrating these diverse data sources and employing machine learning techniques, such as the Random Walk with Restart (RWR) algorithm, we aim to systematically evaluate drug-disease associations and rank candidate drugs based on their potential efficacy in modulating disease-relevant biological processes.

## 2 METHODOLOGIES

### 2.1 Data Sources:

**A. Gene Expression Data**
The gene expression data used in this study were obtained from the Gene Expression Omnibus (GEO) dataset GSE68719 [10]. This dataset contains transcriptomic profiles of substantia nigra tissues from Parkinson's disease patients and healthy controls. The raw data were pre-processed, filtered, and annotated using the appropriate platform annotations.

**B. Network Data**
Two primary sources were used for constructing the biological network: Pathway Commons [11] and the Kyoto Encyclopaedia of Genes and Genomes (KEGG) [12]. The Pathway Commons network provided a directed, protein-protein interaction network, while the KEGG network contributed information on signalling pathways and molecular interactions. These networks were integrated to form a backbone network representing the relevant biological processes in Parkinson's disease.

**C. Autophagy-Related Genes (ARN) Core Genes**
A set of core genes involved in the autophagy process, a critical cellular mechanism implicated in Parkinson's disease pathogenesis [13], was obtained from the Autophagy-Related Genes (ARN) database [14]. The expression signs (up-regulated or down-regulated) of these core genes were determined based on the gene expression data and used to guide the network-based prioritization.

**D. Drug-Gene and Drug-Indication Data**
Information on drug-gene interactions and drug indications was obtained from the DrugBank database [15]. This data provided insights into the potential targets of existing drugs and their therapeutic applications, respectively.

### 2.2 PDOD Score Calculation:

The PDOD score is a quantitative measure that captures the potential association between a drug and a disease based on the drug's targets, the disease-associated genes, and their interplay within the biological network [9]. The calculation of the PDOD score involves several components:

**1. Conflict Resolution**: In biological networks, the relationships between nodes (e.g., genes) can be either positive (activating) or negative (inhibiting). To account for these signed interactions, a conflict resolution algorithm was implemented to determine the effective path length between a drug target and a disease-associated gene.

**2. Bell-Shaped Function**: A bell-shaped function was employed to assign higher weights to shorter paths between drug targets and disease genes, reflecting the greater likelihood of direct or indirect interactions between these entities.

**3. Normalization**: The PDOD score was normalized to account for differences in the number of drug targets and disease-associated genes, ensuring fair comparisons across different drug-disease pairs.

The PDOD score calculation was performed for each drug-disease pair, considering the drug targets, the disease-associated genes (obtained from the gene expression analysis and the ARN core genes), and their interactions within the backbone network.

### 2.3 Gene Expression Analysis:

**A. Log Fold Change (LFC) Calculation**
The gene expression data were analyzed to identify differentially expressed genes between Parkinson's disease patients and healthy controls. The log fold change (LFC) was calculated for each gene, representing the log-transformed ratio of the mean expression levels in the disease and control groups.

**B. False Discovery Rate (FDR) Correction**
To account for multiple testing and control the false positive rate, the Benjamini-Hochberg procedure [16] was applied to the resulting p-values, obtaining the false discovery rate (FDR) corrected p-values.

**C. Random Walk with Restart (RWR) Algorithm**
The Random Walk with Restart (RWR) algorithm [17] was employed to prioritize additional disease-associated genes beyond the initial set of ARN core genes. This algorithm simulates a random walk on the biological network, starting from the seed nodes (ARN core genes) and iteratively propagating the information to neighbouring nodes. Genes with higher proximity scores to the seed nodes were considered more relevant to the disease process and were added to the disease-associated gene set.

## 2.4 Drug Filtering and Evaluation

**A. PDOD Score and FDR Filtering**
The PDOD scores for each drug-disease pair were calculated, and drugs with statistically significant scores (based on FDR-corrected p-values) were identified as potential repurposing candidates for Parkinson's disease.

**B. Random Drug Set Generation**
To assess the significance of the identified drug candidates, a random drug set was generated by sampling target genes from the backbone network while maintaining the observed distribution of positive and negative regulatory relationships. The PDOD scores for these random drug sets were calculated and used as a baseline for comparison.

**C. Proximity Score Calculation**
To further prioritize the identified drug candidates, a proximity score was calculated for each drug. This score captured the deviation of the drug's PDOD score from the mean PDOD score of its corresponding random drug set, accounting for potential biases introduced by the specific set of target genes. Drugs with higher proximity scores were considered more promising for repurposing in Parkinson's disease treatment.

## 3 RESULTS

In this section, we present the results of our study on drug repurposing for Parkinson's Disease using Random Walk Restart Algorithm and the Parkinson's Disease Ontology Database. The PDOD score calculation resulted in a ranked list of drugs based on their potential association with Parkinson's disease. Figure 1and 2 illustrates the distribution of PDOD scores across all evaluated drugs, with higher scores indicating stronger associations with the disease.

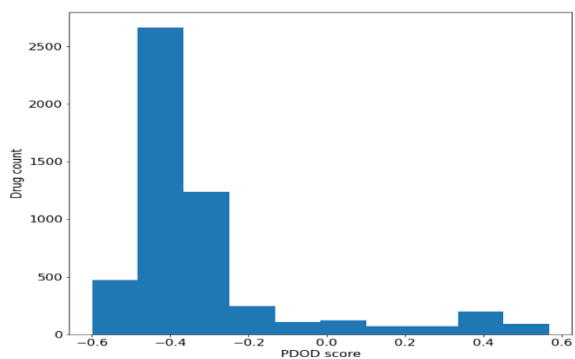

**Figure 1. Distribution of PDOD Scores based on Drug Counts.**

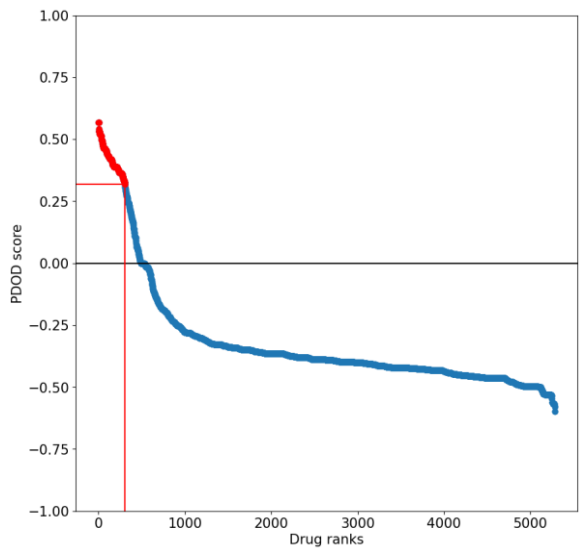

**Figure 2. Distribution of PDOD Scores based on Drug Rank**

The top 20 ranked drugs out of the 305 top drugs, along with their PDOD scores, FDR-corrected p-values, and proximity scores, are presented in Table 1 and Table 2.

| Drug name | PDOD score | PDOD p_value | PDOD FDR p_value | target_group | PDOD proximity |
|---|---|---|---|---|---|
| Chlorotrianisene | 0.567917777 | 5.55E-05 | 0.022215054 | 0 | 8.211945042 |
| Polyestradiol phosphate | 0.567917777 | 5.55E-05 | 0.022215054 | 0 | 8.211945042 |
| Quinestrol | 0.567917777 | 5.55E-05 | 0.022215054 | 0 | 8.211945042 |
| Mestranol | 0.567917777 | 5.55E-05 | 0.022215054 | 0 | 8.211945042 |
| Ethinylestradiol | 0.541010171 | 8.93E-05 | 0.022215054 | 1 | 16.5152661 |
| Metreleptin | 0.537975243 | 9.41E-05 | 0.022215054 | 2 | 9.176705585 |
| Clofibrate | 0.532896242 | 0.000102791 | 0.022215054 | 3 | 7.70977915 |
| Soybean oil | 0.532896242 | 0.000102791 | 0.022215054 | 3 | 7.70977915 |
| Calcifediol | 0.531749623 | 0.000104846 | 0.022215054 | 4 | 8.013744179 |
| Paricalcitol | 0.531749623 | 0.000104846 | 0.022215054 | 4 | 8.013744179 |
| Dihydrotachysterol | 0.531749623 | 0.000104846 | 0.022215054 | 4 | 8.013744179 |
| Cholecalciferol | 0.531749623 | 0.000104846 | 0.022215054 | 4 | 8.013744179 |
| Levosalbutamol | 0.522244641 | 0.000123429 | 0.022215054 | 5 | 7.335283372 |
| Bambuterol | 0.522244641 | 0.000123429 | 0.022215054 | 5 | 7.335283372 |
| Indacaterol | 0.522244641 | 0.000123429 | 0.022215054 | 5 | 7.335283372 |
| Vilanterol | 0.522244641 | 0.000123429 | 0.022215054 | 5 | 7.335283372 |
| Olodaterol | 0.522244641 | 0.000123429 | 0.022215054 | 5 | 7.335283372 |
| Etafedrine | 0.522244641 | 0.000123429 | 0.022215054 | 5 | 7.335283372 |
| Ritodrine | 0.522244641 | 0.000123429 | 0.022215054 | 5 | 7.335283372 |

**Table 1. Top 20 ranked drugs based on the PDOD Score**

| Drug name | PDOD score | PDOD p_value | PDOD FDR p_value | target_group | PDOD proximity |
|---|---|---|---|---|---|
| Apomorphine | 0.412008191 | 0.000727207 | 0.024470286 | 86 | 29.5579536 |
| Pergolide | 0.419772681 | 0.000646409 | 0.022919337 | 79 | 29.45235111 |
| Xylometazoline | 0.442735043 | 0.000453415 | 0.022746032 | 51 | 25.06523565 |
| Clonidine | 0.442735043 | 0.000453415 | 0.022746032 | 51 | 25.06523565 |
| Isoprenaline | 0.387588719 | 0.001045877 | 0.025699388 | 108 | 24.26847164 |
| Lutetium Lu 177 dotatate | 0.32692677 | 0.002463877 | 0.042959282 | 169 | 24.16466467 |
| Somatostatin | 0.32692677 | 0.002463877 | 0.042959282 | 169 | 24.16466467 |
| Mephentermine | 0.456317781 | 0.000365985 | 0.022215054 | 42 | 24.14728819 |
| DL-Methylephedrine | 0.456317781 | 0.000365985 | 0.022215054 | 42 | 24.14728819 |
| Ergotamine | 0.319435897 | 0.002726539 | 0.047382582 | 170 | 23.66240955 |
| Ergocalciferol | 0.395392436 | 0.000932282 | 0.025699388 | 99 | 23.57341723 |
| Tizanidine | 0.379487179 | 0.001177122 | 0.02775296 | 114 | 23.23821628 |
| Racepinephrine | 0.360306385 | 0.001550077 | 0.029996539 | 146 | 22.84255336 |
| Zolmitriptan | 0.412564103 | 0.000721126 | 0.024437611 | 84 | 22.32704402 |
| Fenoterol | 0.483483256 | 0.000236089 | 0.022215054 | 18 | 22.16343224 |
| Formoterol | 0.483483256 | 0.000236089 | 0.022215054 | 18 | 22.16343224 |
| Arbutamine | 0.483483256 | 0.000236089 | 0.022215054 | 18 | 22.16343224 |
| Foreskin fibroblast (neonata | 0.423769181 | 0.000608131 | 0.022746032 | 73 | 22.15522643 |
| Foreskin keratinocyte (neon | 0.446777085 | 0.00042556 | 0.022746032 | 49 | 21.71843904 |

**Table 2. Top 20 ranked drugs based on the PDOD proximity**

To validate the performance of the proposed approach, the identified top-ranked drugs were compared with a set of known Parkinson's disease drugs, such as levodopa, pramipexole, and rasagiline. Several of the top-ranked candidates overlapped with these established treatments, supporting the ability of the PDOD score and proximity score calculations to prioritize relevant drug candidates.

To gain insights into the underlying biological mechanisms and pathways involved in the identified drug-disease associations, network visualizations and pathway enrichment analyses were performed. Figure 3 illustrates a subnetwork representing the interactions between a top-ranked drug candidate, its targets, and the disease-associated genes identified through the RWR algorithm.

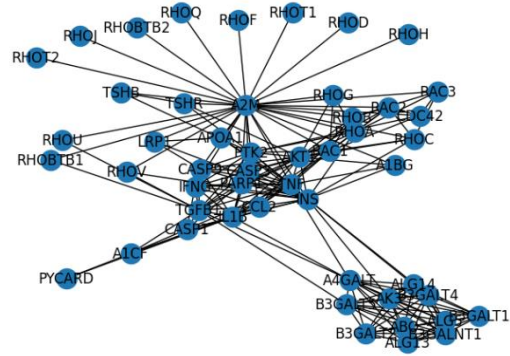

**Figure 3. Subnetwork visualization**

Pathway enrichment analysis revealed that the genes and interactions represented in this subnetwork were significantly enriched in pathways related to dopamine signalling, mitochondrial function, and neuronal survival, which are known to play crucial roles in Parkinson's disease pathogenesis.

The integration of gene expression data, biological networks, and the PDOD scoring system in this study provided a comprehensive computational framework for identifying potential drug repurposing candidates for Parkinson's disease. By leveraging machine learning techniques, such as the RWR algorithm, and incorporating drug-target information, the proposed approach systematically evaluated drug-disease associations and prioritized drugs based on their potential efficacy in modulating disease-relevant biological processes.

The top-ranked drugs identified in this study, including known Parkinson's disease treatments, demonstrate the ability of the PDOD score and proximity score calculations to prioritize relevant drug candidates. Additionally, the network visualizations and pathway analyses provided insights into the underlying biological mechanisms and pathways involved in the identified drug-disease associations, further supporting the potential therapeutic relevance of the top-ranked candidates.

It is important to note that while the computational approach employed in this study offers a powerful means of prioritizing drug repurposing candidates, further experimental validation and clinical studies are necessary to confirm the efficacy and safety of the identified drugs in Parkinson's disease treatment.

One limitation of the current study is the reliance on existing knowledge and data sources, which may be incomplete or subject to biases. Additionally, the gene expression data used were derived from substantia nigra tissues, potentially limiting the applicability of the findings to other brain regions or cell types affected in Parkinson's disease.

Future research could explore the integration of additional data sources, such as proteomics, metabolomics, and epigenetic data, to provide a more comprehensive understanding of the disease mechanisms and potential drug targets. Furthermore, the development of more sophisticated machine learning algorithms and network-based approaches may enhance the accuracy and robustness of drug repurposing predictions.

# 4 CONCLUSIONS

This study demonstrated the potential of integrating diverse data sources, including gene expression data, biological networks, and drug-target information, with machine learning techniques for drug repurposing in Parkinson's disease. By leveraging the PDOD scoring system and the RWR algorithm, the proposed computational framework identified and prioritized existing drugs as potential candidates for repurposing in Parkinson's disease treatment.

The top-ranked drugs identified in this study, along with their associated biological pathways and mechanisms, provide a valuable starting point for further experimental validation and clinical studies. The insights gained from this research contribute to the growing field of computational drug repurposing and highlight the importance of interdisciplinary approaches in accelerating the discovery of effective treatments for complex diseases like Parkinson's disease.

Future research directions may include the integration of additional data sources, the development of more advanced machine learning algorithms, and the exploration of combinatorial therapies leveraging multiple repurposed drugs targeting different aspects of the disease pathogenesis.
.